\documentclass[10pt,letterpaper]{article}
\usepackage{opex3}
\usepackage{amsmath, amsthm, amssymb}
\usepackage{cite}

\newcommand{\be}{\begin{equation}}
\newcommand{\ee}{\end{equation}}
\newcommand{\bea}{\begin{eqnarray}}
\newcommand{\eea}{\end{eqnarray}}

\begin{document}

\title{Frequency modulation spectroscopy at high modulation depth \\ in an indium atomic beam}

\author{N. A. Schine,$^{1,*}$ G. Ranjit,$^{2}$ and P. K. Majumder}
\address{Department of Physics, Williams College, Williamstown, MA 01267}
\address{$^1$Department of Physics, University of Chicago, Chicago, IL 60637$\quad$ \\$^2$Department of Physics, University of Nevada, Reno, NV 89557 $\quad\quad$}
\email{$^*$nschine@uchicago.edu}

\begin{abstract}
We present a detailed analysis of an application of frequency modulation (FM) spectroscopy in the high modulation depth limit. We have recently completed and reported a measurement of the Stark shift in the indium $5p_{1/2}\rightarrow 6s_{1/2}$ 410 nm transition using this spectroscopy method [Ranjit, \emph{et al.} Phys. Rev. A \textbf{87}, 032506 (2013)]. FM spectroscopy proved essential to resolve spectroscopic features in an atomic beam where the optical depth was $\sim 10^{-3}$. A dual-modulation scheme is described which ensures truly background-free FM signals even in the low-density limit.  Lock-in detection of the FM signal was accomplished using both the fundamental (1f) and second-harmonic (2f) modulation frequencies.  A derivation of both the 1f and 2f signal line shapes in the high-modulation-depth limit is presented.  These line shapes form the basis for quantitative fits to a wide variety of experimental FM spectra.
\end{abstract}


\section{Introduction}

Frequency modulation techniques have been used as an experimental tool for small signal detection in a broad range of disciplines. For example, FM techniques are an important feature of current cantilever atom-force-microscopy experiments\cite{Albrecht91}. Here we focus on a laser spectroscopic application of frequency modulation, first introduced several decades ago as a sensitive measurement technique for both atomic and condensed matter physics applications\cite{Bjorklund80,Bjorklund81,BjorkFM}. FM atomic spectroscopy has included measurement of small absorption and/or dispersion signals in circumstances for which the light is treated in both a classical and quantum mechanical fashion \cite{Grimm89,Sansonetti95,Vladimirova09}. FM spectroscopy has also been used in condensed matter systems to explore various perturbations to solid crystal structures \cite{Moerner86,Moerner89}. All of these applications use first harmonic demodulation (1f), in which the transmission signal is demodulated at the same frequency as that at which the laser is modulated. While less common, second harmonic (2f) demodulation has been employed in analyses of molecular vapors that have complicated spectra \cite{Bomse92,Webster87,Sachse87,Loewenstein88,Liu04}. Additionally, most applications use relatively low modulation depth, where the optical spectrum  is well approximated by considering the carrier frequency and  first-order sidebands only.  An example of work at higher modulation depth is described in \cite{Vladimirova09,Liu04} where the authors employed direct modulation of the laser diode current to produce multiple side bands. Such implementation can achieve modulation depths, $\beta$, of up to $\beta \simeq 10$, but at the cost of significant residual amplitude modulation (RAM), resulting in complications in the subsequent analysis.

We have recently completed a measurement of the Stark shift in the indium $5p_{1/2}\rightarrow6s_{1/2}$ transition which utilized high-modulation-depth frequency-modulation spectroscopy with both first and second harmonic demodulation \cite{Ranjit13}. We implemented the frequency modulation with an electro-optic modulator (EOM), which introduces negligible RAM and can produce modulation depths well above unity in the optical spectrum. We employ high modulation depths to increase signal size and make use of multiple sidebands for frequency scale linearization and calibration, as described below. 

We first present a brief overview of the experimental setup and the context of this work. We then present the theoretical description of both the 1f and 2f demodulated FM profile for arbitrary modulation depths which we have used to create a line shape model to fit experimental spectra. Finally, we identify and discuss several non-idealities  in the experimental FM procedure which we had to address in order to complete our analysis of the spectra.

\section{Background and experimental details}

\subsection{Testing atomic theory}

The Stark shift is the change in the frequency of an atomic transition due to the presence of an electric field. By measuring this shift for a known electric field, we can compare our result to that of \emph{ab initio} atomic theory calculations which model the multi-electron Group III atomic systems which we study\cite{Saf09, Saf13}. In the one-electron central field (OECF) approximation, second order perturbation theory predicts that a static electric field causes an energy shift, $\Delta W$, to each energy level, $j$, of the valence electron that is quadratic in the magnitude of the static electric field, $E$, i.e. $\Delta W_j = -\frac{1}{2}\alpha_{0(j)}E^2$. The polarizability of the state, $\alpha_{0(j)}$, characterizes the magnitude of the Stark shift for a given electric field.  Thus for our optical transition, we measure a frequency shift that is quadratic in the field strength and proportional to the difference in polarizabilities of the final and initial states, $\Delta \alpha_0$.  In \cite{Ranjit13}, we described in detail an experiment which produced the result $\Delta \alpha_0 = 1000.2(2.7)$ $a_0^3$ (in atomic units) for the particular indium transition under study.  The precision of this result represented a 30-fold improvement over previous work, and was a direct result of the FM spectroscopy technique discussed here.  An independent, \emph{ab initio} theoretical calculation of  this quantity undertaken  at the same time as  our experimental work\cite{Saf13} produced the value $\Delta \alpha_0 = 995(20) a_0^3$, in excellent agreement with our experimental result.

\subsection{Experimental apparatus and Stark shift data collection}

\begin{figure}[t]
\centering
\includegraphics[width = .6\textwidth]{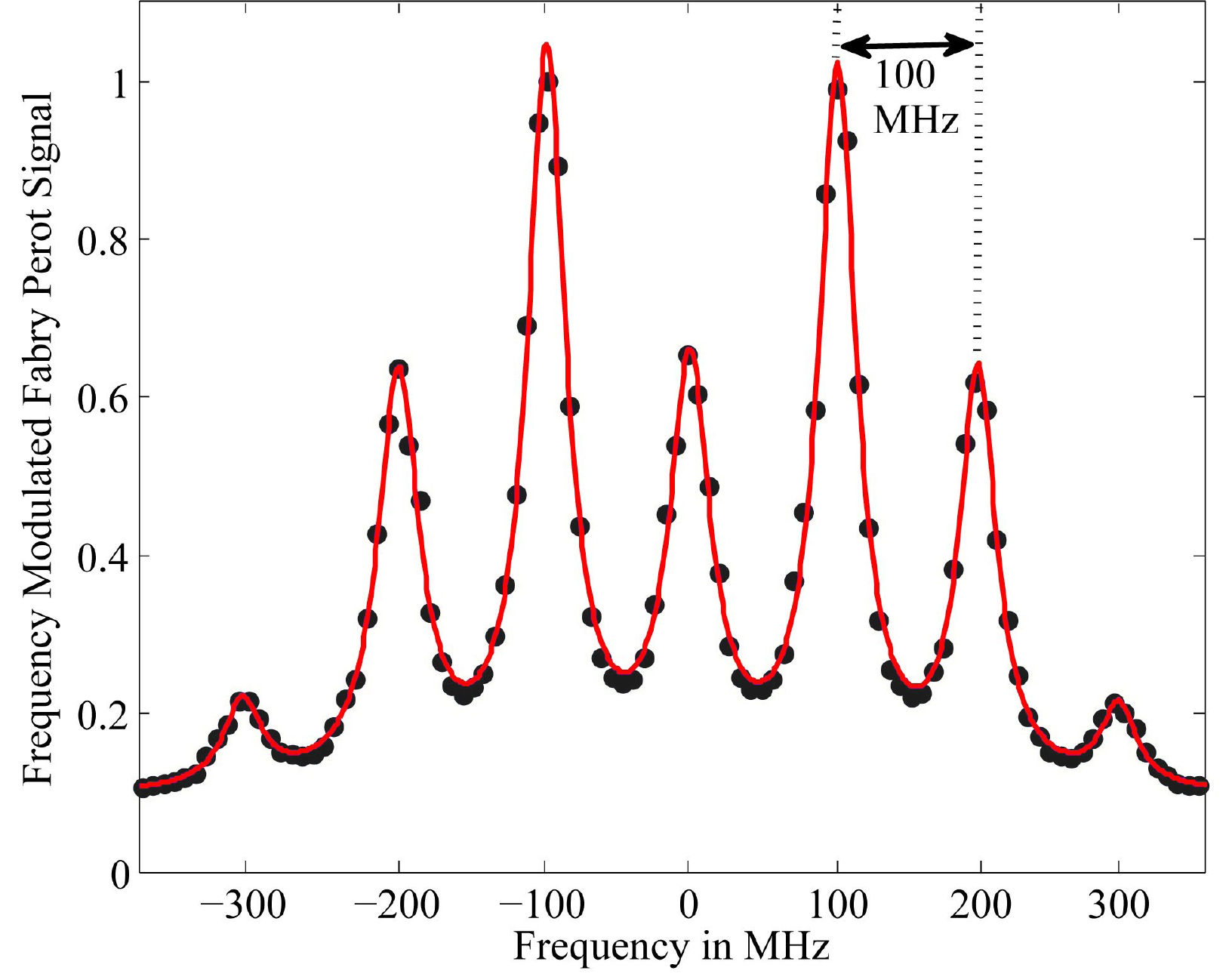}
\caption{An experimental Fabry-P\'erot transmission spectrum of FM laser ($\omega_m=100$ MHz) at moderate modulation depth ($\beta \simeq 2$), showing three sidebands on each side of the carrier (central peak at 0 MHz). This spectrum was obtained by scanning the laser over a single longitudinal mode of the Fabry-P\'erot.  The solid line is a fit to a sum of seven Lorentzians.}
\label{fig:FMFP}
\end{figure}

We now briefly introduce the experimental setup with which we performed FM spectroscopy. Further experimental details can be found in \cite{Ranjit13}.  A commercial Littrow-configured external cavity GaN laser diode system (Toptica Photonics, DL 100) produces approximately 10 mW of light at 410.2 nm. We pass the light through an optical isolator and then frequency modulate the laser using an EOM (New Focus Model 4001). The combination of a 100 MHz synthesizer (Hewlett-Packard 8656B) and power amplifier  (Mini-Circuits ZHL-03-5WF) deliver several Watts of power to the resonant cavity of the EOM.  The range of RF power we used produced modulation depths between roughly 2 and 4.  A portion of the modulated blue laser light was directed into a  confocal Fabry-P\'erot interferometer (Burleigh RC-110) with finesse of roughly 40 and a free spectral range (FSR) of 1 GHz. We passively stabilize the Fabry-P\'erot (FP) cavity by surrounding it with acoustical insulation, minimizing thermal and vibrational noise. When the center frequency of the laser is scanned, the FP transmission signal reveals the sideband spectrum of the modulated laser.  Since for our case the sideband amplitude was negligible after fourth-order, the FP transmission spectrum is not complicated by overlapping peaks from subsequent longitudinal modes of the cavity.  Moreover, since each sideband peak is precisely 100 MHz apart, we used this spectrum in our experiment to linearize the diode laser frequency scan, and to provide a reliable absolute frequency scale for our measurements of atomic spectral shifts, which were collected simultantaneously. Figure \ref{fig:FMFP} shows a typical FP transmission scan.   Additionally, we collected the saturated absorption signal from a supplementary indium vapor cell to provide a frequency reference with which to compare electric field on and off scans.
 The primary laser beam is directed through our atomic beam apparatus via anti-reflection coated windows. The vacuum chamber housing the atomic beam is maintained at a pressure of roughly 10$^{-6}$ torr. The source chamber houses a custom-designed oven that heats approximately 100 g of indium to 900$^o$ C. Indium vapor effuses out of  a multi-slit faceplate towards the interaction region 40 cm away. Multiple collimators produce an atomic beam that is  1.5-cm-wide by 2-mm-high. This beam passes through an in-vacuum chopping wheel assembly, and finally passes between a pair of polished stainless steel plates with 1 cm separation to  which we apply a potential difference in the range of 10 - 20 kV. The frequency modulated 410 nm laser beam passes between the plates and intersects the atomic beam at right angles at the center of the plate assembly. The transmitted light is subsequently collected outside the vacuum apparatus  by a fast 1 GHz bandwidth photodiode (New Focus Model 1601). This photodiode signal is demodulated by a RF lock-in amplifier (SRS model SR844) that is referenced to the 100 MHz signal from our synthesizer. As is discussed in detail below, at our very low optical depths, the RF-demodulated signal contains a small frequency-dependent background signal of purely optical origin. We eliminate this non-atomic background completely by performing a second low-frequency demodulation of the laser transmission signal using the 500 Hz reference signal of the generator which drives the in-vacuum atomic beam chopping wheel. Thus we direct the output of the SR844 to a second lock-in amplifier (SR810)  which produces the background-free atomic signal.

We took laser scans with an alternating on-off sequence of high voltage. Over the course of one month, we obtained several thousand individual Stark shift measurements, distributed over high voltage values between 10 kV and 20 kV, and taken over a wide range of experimental and measurement conditions. For each pair of field on / field off scans, a data analysis program written in Matlab first performed the frequency scale linearization and calibration by fitting the FP spectra to a sum of Lorentzians.  Next, the supplemental saturated absorption spectra were analyzed to produce a common reference frequency point.  Finally we performed fits to the FM-demodulated spectra, as discussed below, to determine the Stark shift.

\section{FM spectroscopic line shapes}

To begin, we model the electric field of the frequency modulated laser in standard fashion as,
\be
E(t) = E_0 \sum_{n=-\infty}^{\infty} J_n(\beta) e^{i(\omega_c+n\omega_m)t},
\label{Eqn:FM2}
\ee 
where $\omega_c$ and $\omega_m$ are the carrier and modulation frequency respectively.  When the laser passes through a sample of atoms, the field is modified by a complex transmission function, $T(\omega) = e^{-\delta(\omega)-i\phi(\omega)}$, where $\delta(\omega)$ and $\phi(\omega)$ are the absorption and dispersion profiles respectively. In general, when both Lorentzian (e.g. natural lifetime) broadening and Gaussian (e.g. Doppler) broadening are significant, these profiles are given by a convolution of Lorentzian and Gaussian known as a Voigt profile. 

To make our line shape fitting procedure more efficient, we choose to employ an analytic approximation to the Voigt profile rather than repeatedly calculating computationally expensive convolution integrals.  This approximation \cite{vetterth, ChiRei68} involves an infinite sum which can be truncated (in our case after only a few terms), and an adjustable parameter (`$h$' in Eq. \eqref{Eqn:ChiReichAlg}). We first define two unitless quantities upon which the profiles depend: $x$, the ratio of Lorentzian to Doppler component widths, and $a$, the ratio of the frequency detuning from resonance to the Doppler width.  Then, in terms of the complex quantity $z \equiv x - i a$, we have

\bea
\delta(x, a) + i \phi(x, a)  \approx  
 \frac{ha}{\sqrt{\pi} z} + \frac{\sqrt{\pi}\,2a\,e^{z^2}}{1-e^{2\pi z /h}} + \frac{2haz}{\sqrt{\pi}}\sum_{n=1}^{N}\frac{e^{-n^2h^2}}{n^2h^2+z^2} &.
\label{Eqn:ChiReichAlg}
\eea

One can find the value of the parameter $h$ which optimizes the agreement of the analytic approximation with the exact Voigt profile for a given ratio of component widths. For our case, where the ratio of Gaussian to Lorentzian components is roughly 3:1, we find that using $h = 4.01$, and employing five terms in the summation produces profiles in excellent agreement with exact numerical counterparts. At that point, the residual point-to-point fluctuations in our experimental spectra exceeds any residual imperfections in the fit model.  Also, in our application we ultimately measure the frequency shift of identically-shaped spectra upon application of a large static electric field, so that small inaccuracies in the fit would affect both spectra equally, and not influence our determination of the shift.

Having defined $\delta(\omega)$ and $\phi(\omega)$, we now write down the electric field of frequency modulated laser after interacting with the atoms in terms of the transmission function $T(\omega)$.
\be
E_T(t) = E_0 e^{i\omega_c t}\sum_{n=-\infty}^{\infty} J_n T_n e^{in\omega_m t},
\label{eqn:TransmittedEfield}
\ee
where $J_n \equiv J_n(\beta), \,\, T_n \equiv e^{-\delta_n - i \phi_n} \approx 1 - \delta_n - i \phi_n, \,\, \delta_n \equiv \delta(\omega_c + n \omega_m), \text{ and} \,\, \phi_n \equiv \phi(\omega_c + n \omega_m).$ Our experimental conditions, with optical depths of  roughly $10^{-3}$,  justifies the expansion of the transmission function for small absorption and dispersion.

Multiplying by the complex conjugate provides terms that oscillate in time at non-negative integer multiples of the modulation frequency. A slow photodiode will average out all time varying terms, revealing only the DC term,
\be
I_{DC} =  I_0 \sum_{n=-\infty}^{\infty} J_n^2 e^{-2\delta_n} = E_0^2 \sum_{n=-\infty}^{\infty} J_n^2 (1-2\delta_n)
\ee
In our application, the optical depth is sufficiently small that such direct observation of the absorption is impossible.

\subsection{First Harmonic Demodulation}

\begin{figure*}[t]
\includegraphics[width=1\textwidth]{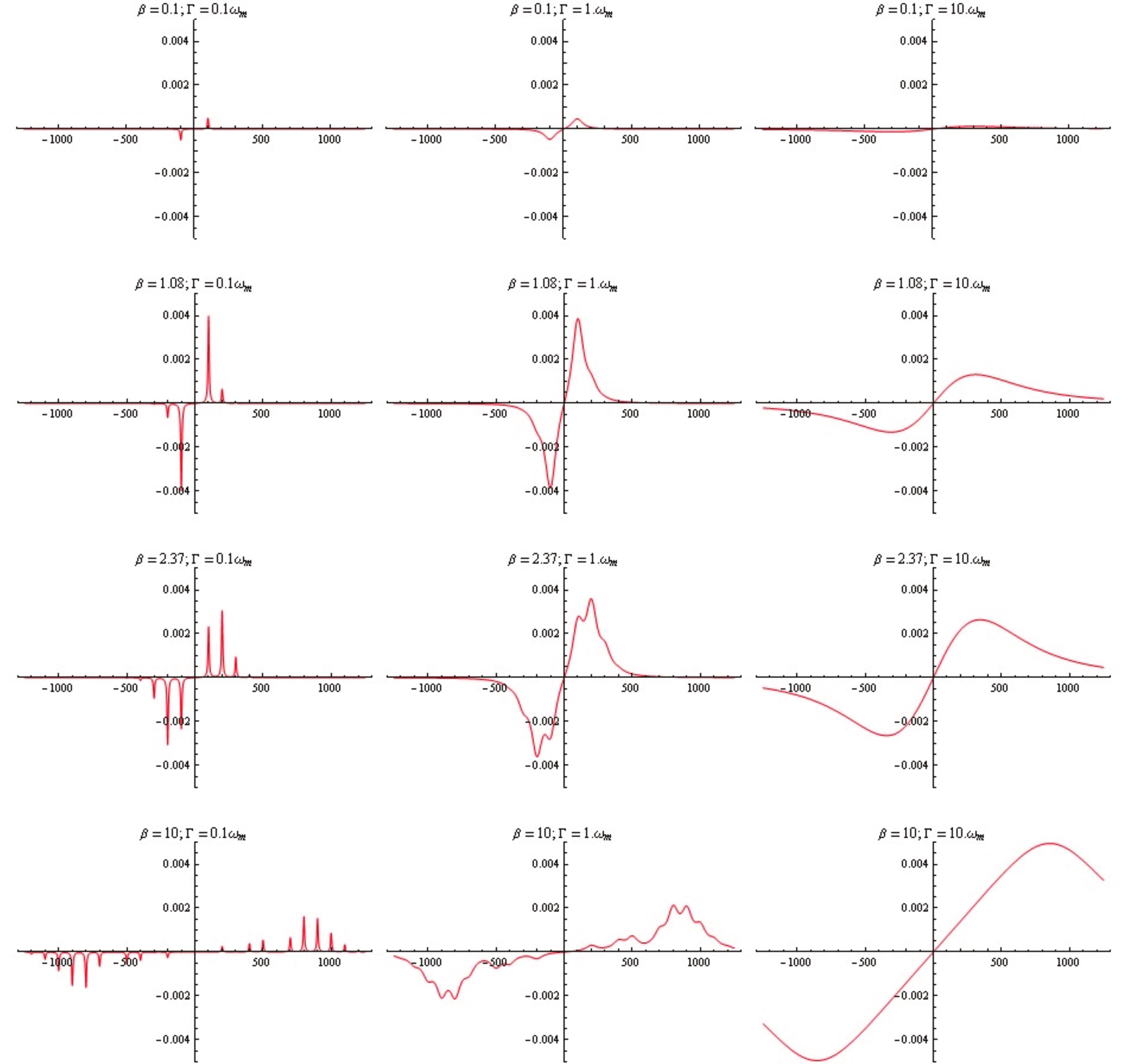}
\caption{Simulated in-phase ($\theta_d = 0$) demodulated FM transmission signals for a variety of modulation depths, $\beta$, and  spectral widths, $\Gamma$, with $\omega_m=100 MHz$. For simplicity, we take the spectral broadening in these simulations to be purely Lorentzian. Note that all 1f-demodulated line shapes are odd about the profile center. Our typical range of experimental parameters produce spectra that closely resemble those in the middle of this array of plots. Horizontal axes are MHz in all cases.  Vertical axes (arbitrary units) are kept consistent for ease of comparison. }\label{Fig:SimParameterSpace}
\end{figure*}

To take advantage of the FM spectroscopy technique, we therefore make use of a lock-in amplifier that picks out terms in the transmitted signal which oscillate at $\omega_m$ (or $2\omega_m$). We derive these profiles below, assuming that the absorption and dispersion are small, slowly varying functions, namely $|\delta_0 - \delta_n| \ll 1$ and $|\phi_0 - \phi_{n}| \ll 1$, but placing no restriction on the modulation depth, $\beta$. For first-harmonic detection, relevant terms in the transmitted intensity expression are those corresponding to sidebands whose indices differ by one, and therefore oscillate at $\omega_m$. Appropriate manipulation of summation indices then leaves us with
\be
|\tilde{E}_T(t)|^2_{1f} = E_0^2 \sum_{n=-\infty}^{\infty}J_nJ_{n+1} \left( T_{n+1}^* T_n e^{-i\omega_m t} +  T_n^* T_{n+1} e^{i\omega_m t}\right).
\ee
Recasting this as a sum over non-negative indices, we then have
\be
|\tilde{E}_T(t)|^2_{1f} = E_0^2 \sum_{n=0}^{\infty}J_n J_{n+1} \Bigl[ (T_{n+1}^* T_n - T_{-n}^*T_{-n-1}) e^{-i\omega_m t} + (T_n^*T_{n+1}-T_{-n-1}^*T_{-n} ) e^{i\omega_m t} \Bigr].
\label{Eqn:1f-1}
\ee

Finally, making use of the assumption that the absorption is small to rewrite these products of transmission functions, we separate the expression into sines and cosine terms and arrive at the 1f demodulated signal, with $\theta_d$ signifying the relative phase between the detected signal and the lock-in amplifier reference signal.
\bea
I_{demod}^{1f}(t)  =    I_0 e^{-2\delta_0} \sum_{n=0}^{\infty}J_n J_{n+1} \Bigl[ \left(\delta_{-n} - \delta_n +\delta_{-n-1}-\delta_{n+1} \right)\cos(\theta_d)  
  + \Bigl. \nonumber \\
 \Bigr. \left( \phi_{n+1}-\phi_n+\phi_{-n-1}-\phi_{-n}\right) \sin(\theta_d) \Bigr]. 
\label{Eqn:1fDemod}
\eea
This expression reproduces the results of \cite{Supplee94}.  Figure \ref{Fig:SimParameterSpace} explores the profiles defined by Eq. \eqref{Eqn:1fDemod} over a wide range of modulation depths, $\beta$, and for various values of the ratio of spectral line width (assumed here to be purely Lorenztian) to modulation frequency. In the narrow spectral feature regime, the peak associated with each sideband is independent of its neighbors, while in the wide spectral feature regime, all peaks overlap to produce a simpler composite profile. In our experimental regime, the spectral line-width is comparable to the modulation frequency, producing profiles with multiple partially-overlapping peaks. Additionally, Fig. \ref{Fig:SimParameterSpace} demonstrates the efficacy of moving to the moderate modulation depth regime ($\beta \gtrsim 1$), as the signal size in that case is roughly an order of magnitude greater than in the low modulation depth limit. This increase in sensitivity was of great importance in our high-precision experiments. We note that for situations where spectral width and modulation frequency are comparable, there is no further gain in signal size in the very high modulation depth regime, as laser power becomes spread over many sidebands.

\begin{figure}[h!]
\vspace{6pt}
\centering
\includegraphics[width=.95\textwidth]{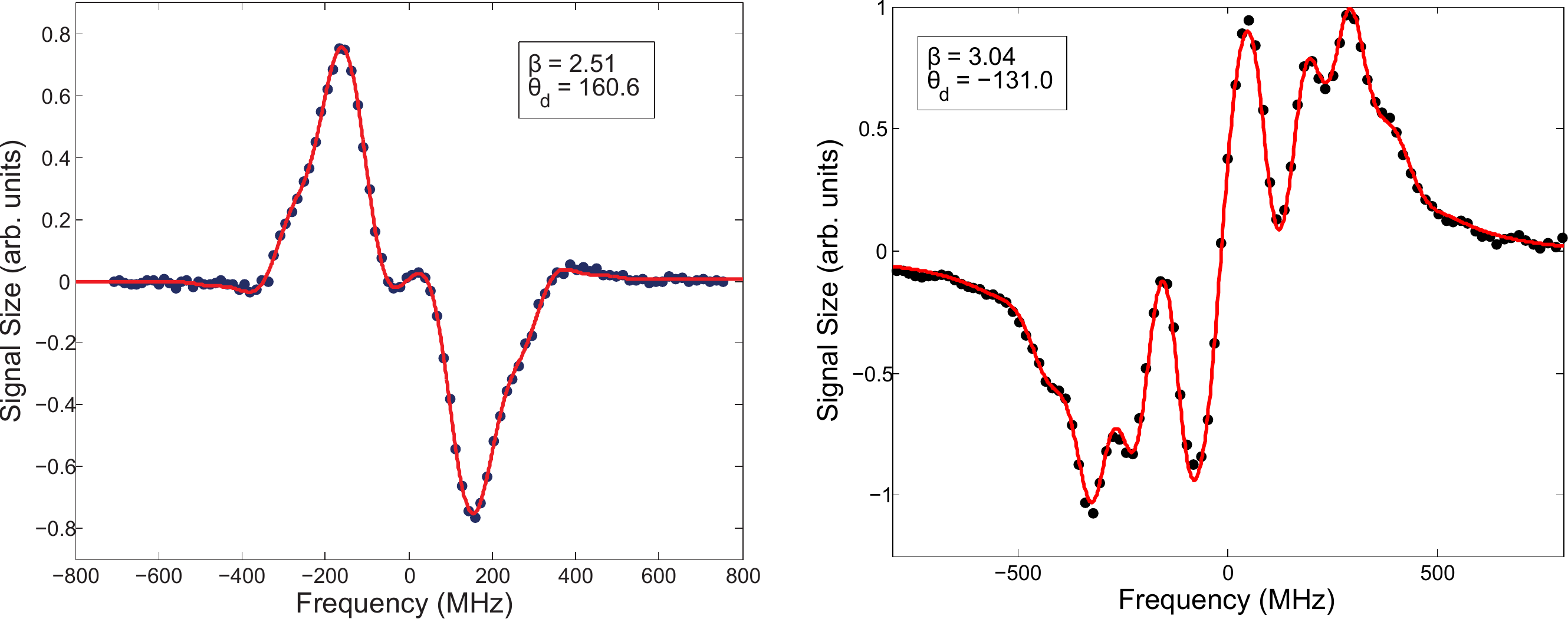}
\caption[2f FM Signals]{Two representative 1f experimental FM spectra fit using the full analytic form given by Eq. \eqref{Eqn:2fSig} (solid line). Fitted values for modulation depth and detection phase angle  (degrees) are listed on each plot.  }
\label{Fig:FM1fSignals}
\vspace{-10pt}
\end{figure}

We use the analytic expression of Eq. \eqref{Eqn:1fDemod} as the basis for our fits to experimental 1f spectra.  Our line shape model includes as fit parameters the center frequency, overall amplitude, optical depth $\beta$, relative phase $\theta_d$, and the component Lorentzian and Gaussian widths.  A representative set of repeated scans at nominally identical operating conditions produces consistent values for all parameters at the 1\% level or below.  Moreover, all of our fit results produced Lorentzian width values near 22 MHz, in excellent agreement with the natural line width of the indium $6s_{1/2} - 5p_{1/2}$ transition.  The fitted value of the Gaussian width was 60(1) MHz, which is consistent with the residual Doppler broadening expected given our atomic beam collimation geometry. We found that the fits improved if we allowed the relative amplitudes of the sidebands to vary slightly from those predicted by the single modulation depth parameter.  These additional amplitude correction factors (assumed to be the same for corresponding positive and negative-order sidebands)  generally ranged from 0.85 to 1.15.  That is, the deviation from `ideal' sideband amplitude distribution was typically less than 15\%.  In section \ref{sect:incompleteMod} we discuss a possible model for this behavior resulting from spatial variation of frequency modulation strength across our finite laser beam as it passes through the resonant cavity within the EOM.  We show in Fig. \ref{Fig:FM1fSignals} a representative set of two 1f-demodulated experimental spectra, taken over a range of operating conditions.  Fitted values for the modulation depth and phase angle are listed in each case.

\subsection{Second Harmonic Demodulation}
Conveniently, our RF lock-in amplifier can be easily  configured to demodulate at the second harmonic of the modulation frequency, in this case at $2\omega_m = 200$ MHz. In contrast to the 1f-demodulated signals, 2f demodulation results in line shapes which are even about line center. In our experimental application, changing the line shape in this way and utilizing these distinct spectral shapes required the use of independent fitting programs, therefore acting as a useful check on the reliability of the fitting programs and our overall experimental method. Ultimately the excellent agreement we found between analysis of experimental spectra in the 1f vs. 2f configuration helped to minimize concerns about systematic errors that often plague precision measurements\cite{Ranjit13}.  Here we go into further detail deriving the complete 2f-demodulated signal, as it has not to our knowledge been previously worked out for the case of large modulation depth.

We start again with the electric field transmitted through the atomic sample as given by Eq. \eqref{eqn:TransmittedEfield}. The transmitted intensity is again given by a product of sums. A lock-in amplifier configured for 2f detection picks out those terms in the intensity that oscillate at $2\omega_m$. Taking the indices of the sums to be $n$ and $l$, then for each value of $n$, we obtain the desired terms when $l=n-2$ or if $l=n+2$, that is, 
\bea
|\tilde{E}_T(t)|_{2f}^2 &=& E_0^2 \Bigl[ \sum_{n=-\infty}^{\infty} \left( J_n T_n^* e^{-in\omega_m t}\,J_{n-2}T_{n-2}\,e^{i(n-2)\omega_m t} \right) + \Bigr. \nonumber \\
&& \quad \quad \quad \quad \quad \quad \quad \quad \quad \quad \Bigl. \sum_{n=-\infty}^{\infty} \left( J_n T_n^* e^{-in\omega_m t}\,J_{n+2}T_{n+2}\,e^{i(n+2)\omega_m t} \right) \Bigr] \nonumber \\
&=& E_0^2 \sum_{n=-\infty}^{\infty} J_n J_{n+2}\left(T_{n+2}^*T_n\,e^{-2i\omega_m t} + T_n^*T_{n+2}\,e^{2i\omega_m t} \right). 
\eea
We redefine indices to be non-negative and utilize the property of Bessel functions, $J_{-n} = (-1)^n J_n$, and find that
\bea
&&|\tilde{E}_T(t)|_{2f}^2 = E_0^2 \Bigg[-J_1^2 \left( T_1^*T_{-1}\, e^{-2i\omega_m t} + T_{-1}^*T_{1} \, e^{2i\omega_m t} \right) + \cdots \nonumber \\
&&\sum_{n=0}^{\infty} J_n J_{n+2} \Bigl( \left(T_{-n}^*T_{-n-2}+T_{n+2}^*T_n \right) \,e^{-2i\omega_m t} + \left(T_{-n-2}^*T_{-n} +  T_n^*T_{n+2}\right)\,e^{2i\omega_m t} \Bigr) \Biggr].
\eea
We can now combine and re-express combinations of transmission functions.  For example
\be
T_{-n}^*T_{-n-2}+T_{n+2}^*T_{n} = e^{-2\delta_0}[ \left( 2+4\delta_0 -\delta_n - \delta_{-n} - \delta_{n+2} - \delta_{-n-2} \right) +  i\left( \phi_{-n} - \phi_{n} + \phi_{n+2} - \phi_{-n-2} \right) ].
\ee
At this point, we can expand the complex exponentials into sines and cosines and combine terms to collect the in-phase  and the quadrature components. Finally, the output of the lock-in amplifier with reference frequency $2\omega_m$ and demodulation phase $\theta_d$ is found to be
\bea
I_{demod}^{2f} = I_0 e^{-2\delta_0} \Biggl[ \Bigl(-J_1^2 (1+2\delta_0 -\delta_1-\delta_{-1}) &+& \Bigr.\Biggr.\nonumber \\
&&\hspace{-80pt}\Bigl. \sum_{n=0}^{\infty} J_n J_{n+2} (2+4\delta_0 -\delta_n - \delta_{-n} - \delta_{n+2} - \delta_{-n-2}) \Bigr) \cos(\theta_d) + \nonumber \\
&&\hspace{-155pt}\Bigl( -J_1^2(\phi_1-\phi_{-1}) + \sum_{n=0}^{\infty} J_n J_{n+2} (\phi_{-n} - \phi_{n} + \phi_{n+2} - \phi_{-n-2}) \Bigr) \sin(\theta_d) \Biggr].
\label{Eqn:2fSig}
\eea

\begin{figure}[h!]
\centering
\includegraphics[width=.95\textwidth]{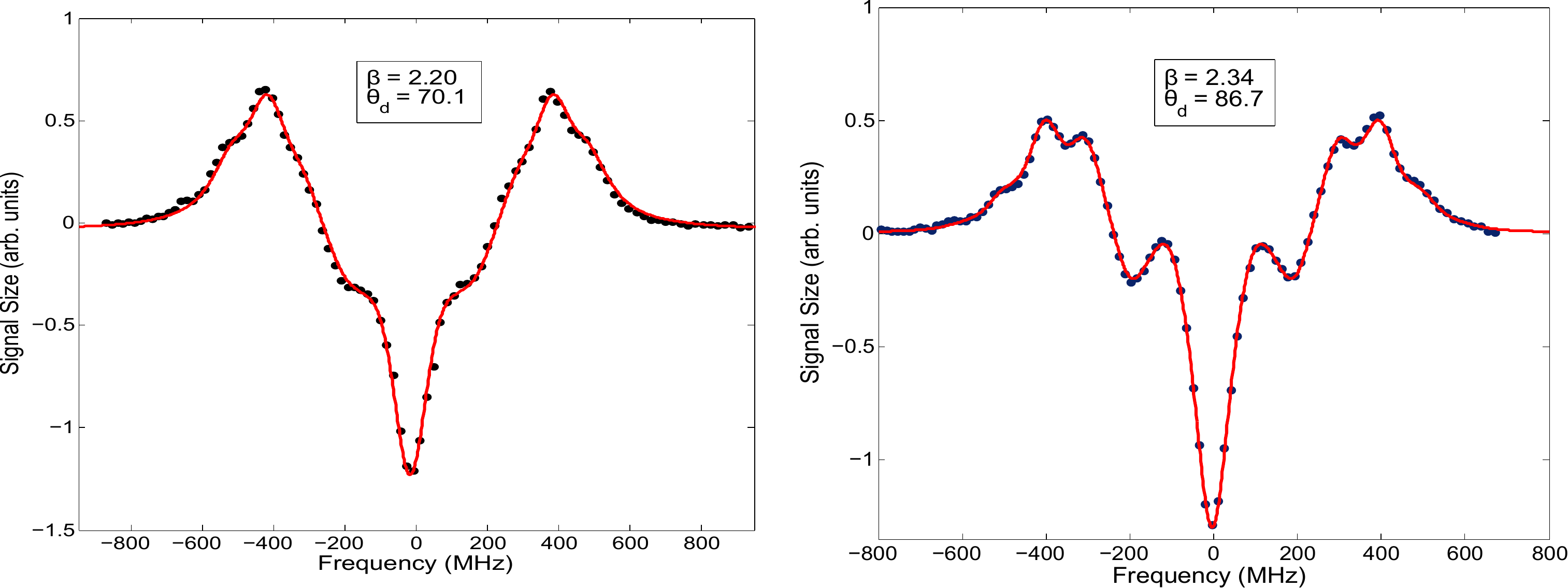}
\caption[2f FM Signals]{Two representative 2f experimental FM spectra fit using the full analytic form given by Eq. \eqref{Eqn:2fSig} (solid line). Fitted values for modulation depth and detection phase angle (degrees) are listed on each plot.  }
\label{Fig:FM12Signals}
\end{figure}

Figure \ref{Fig:FM12Signals} shows a pair of 2f-demodulated experimental spectra fit to the line shape model in Eq. \eqref{Eqn:2fSig}.  In our Stark shift experiment, where we require only a reliable measure of the spectral \emph{shift} between two nearly identical line shapes, it is of course possible to analyze the complex spectra shown here with simplified models involving polynomials or Lorentzian functions, that focus only on the central feature of each spectrum.  As discussed in \cite{Ranjit13}, we also pursued this simplified analysis to obtain line centers and frequency shifts.  Once again, the excellent agreement between results from the simplified models and those from the exact FM line shape analysis provided confidence in the reliability of our final experimental result.

\section{Non-idealities in the observed FM line shapes}

We conclude with a brief discussion of the ways in which our frequency modulation line shapes deviate in practice from the idealized model represented by Eq. \eqref{Eqn:FM2}. 

\subsection{Residual optical background}

In an independent set of preliminary studies, we were able to produce and analyze FM transmission signals in a heated indium vapor cell whose temperature could be controlled.  By systematically reducing the atomic density, we observed that for very low optical depths (OD $<$ .01) we could observe, superimposed on the (Doppler-broadened) atomic FM signal, a small frequency-dependent background pattern that remained consistent in size and shape over time scales of minutes to hours.  Realignment of the laser beam through the EOM caused this background pattern  to change size and shape.  Indeed,  in our ultimate atomic beam application where the low atomic density produced OD $\sim 10^{-3}$, such background patterns were also observed. If not addressed, this feature can lead to inaccuracies in the FM line shape fits.  Of particular concern were possible mechanical drifts or variations in laser beam pointing that could \emph{alter} this background pattern on time scales similar to our Stark shift field-switching times, in which case these background features could cause systematic errors in our frequency shift determinations.   Although the faces of our EOM are anti-reflection coated, small internal reflections within the EOM can presumably cause an etaloning effect, which would add a small frequency-dependent transmission component.  We attribute our background pattern to this mechanism. Because the residual reflection is small, this mimics the effect of a very low-finesse optical cavity producing small sinusoidal signal variation rather than sharp transmission peaks.  It is not surprising, then, that realignment of the laser beam through the EOM can change the phase and amplitude of this background pattern.
 
Regardless of the exact origin of the background, we were able to completely eliminate it by modulating the atomic beam with our in-vacuum chopping wheel.  By superimposing the low-frequency modulation on the RF modulation created by the EOM, we can use a second low-frequency lock-in amplifier to subtract the purely optical background signal, leaving us with the background-free atomic spectrum.  At the time scale of this chopping wheel rotation (few hundred Hz) we find that the optical background pattern remains stable, allowing effective subtraction, as can be seen in the variety of experimental spectra included here.

\subsection{Residual amplitude modulation}

\begin{figure}[h!]
\centering
\includegraphics[width=.50\textwidth]{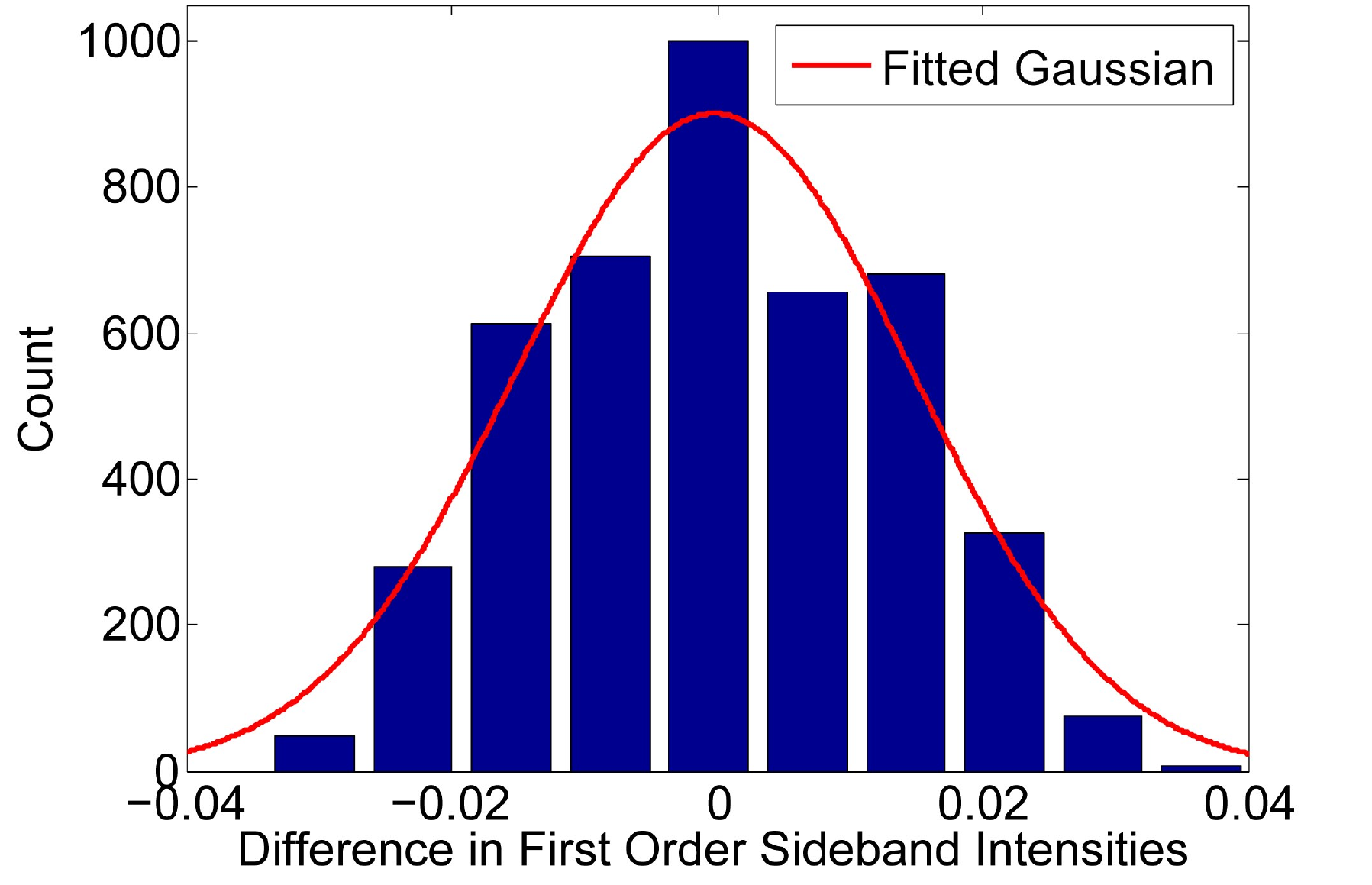}
\caption{A histogram of the differences between the positive and negative first-order sidebands as measured from several thousand FP transmission spectra taken under a variety of experimental conditions. The average value for the difference is 0.00(1), implying that any residual RAM is below the 1\% level.}
\label{fig:D1Order}
\end{figure}

All phase modulators such as our EOM have the potential to cause residual amplitude modulation at $\omega_m$ in addition to the desired frequency modulation. RAM adds additional in-phase frequency sidebands at  $\pm \omega_m$. Since true FM produces first-order sidebands with opposite phase, RAM causes a net imbalance in the magnitude of the two first-order sidebands, and overall asymmetry in the sideband distribution. Also, RAM is directly proportional to the input laser power. Therefore, while laser intensity noise is effectively filtered from the purely FM signal, the signal quality of the demodulated line shape may be compromised by such intensity noise in the presence of RAM\cite{BjorkFM}. We performed a comprehensive  analysis of the peak heights in a large number of our Fabry-P\'erot transmission scans (Fig. \ref{fig:FMFP}), comparing the intensities of the two first-order sidebands. The results, shown in Fig. \ref{fig:D1Order}, show that any RAM is well below the level of 1\%, and thus has negligible impact on our line shapes and analysis.  We note that direct current modulation of the laser diode, while a simpler and inexpensive alternative, typical results in much larger RAM, especially in the high-modulation-depth limit in which we worked.

\subsection{Distributed Modulation}
\label{sect:incompleteMod}

\begin{figure}[h!]
\centering
\includegraphics[width=.5\textwidth]{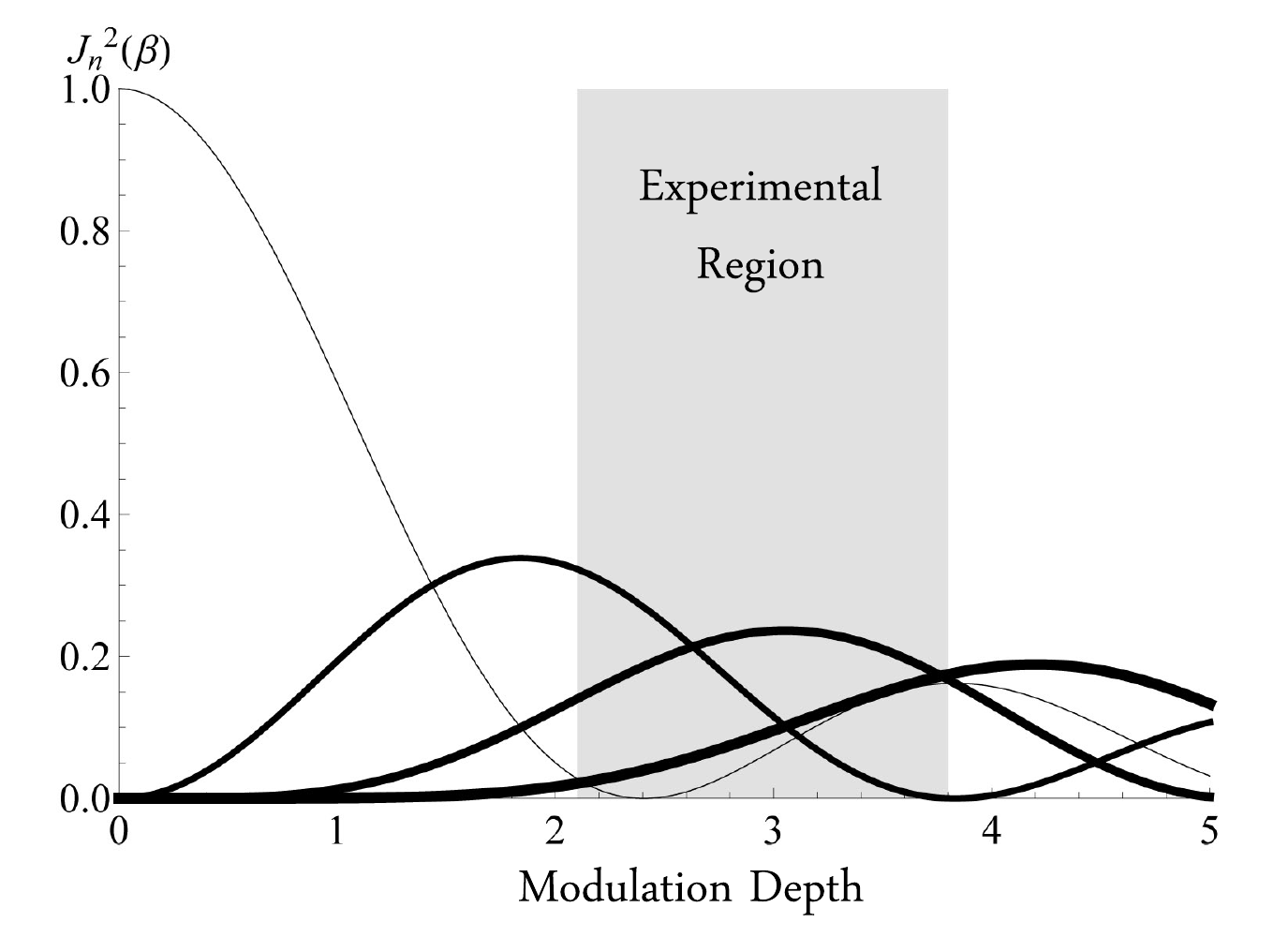}
\caption[Bessel Functions]{The squares of the first four Bessel functions $J_n^2$ for $n=0,1,2,$ and $3$ plotted with lines of increasing thickness for higher-order functions. The range of modulation depth values utilized experimentally is indicated by the shaded band. The lower bound of the experimental region is determined by the minimum $\beta$ at which the third order side band is easily measurable.}
\label{Fig:BesselFcns}
\end{figure}

Finally, we discuss a simple physical model which can help explain why our observed sideband intensity distribution does not precisely match the prediction of FM theory. Evidence to support this model is provided, again, by the measurement  of relative heights of sideband peaks in the FP transmission spectrum when we scan the laser over a single longitudinal mode. FM theory would predict that for modulation depth, $\beta$, the peak height of the $n^{th}$ sideband in the laser spectrum to be proportional to $J_n^2(\beta)$, where $J_n$ is the $n^{th}$-order Bessel function.  Figure \ref{Fig:BesselFcns} plots the first few of these functions over a range of modulation depths. In particular, this model predicts that each sideband component should go to zero for a particular modulation depth. 

\begin{figure}[h!]
\centering
\includegraphics[width=.5\textwidth]{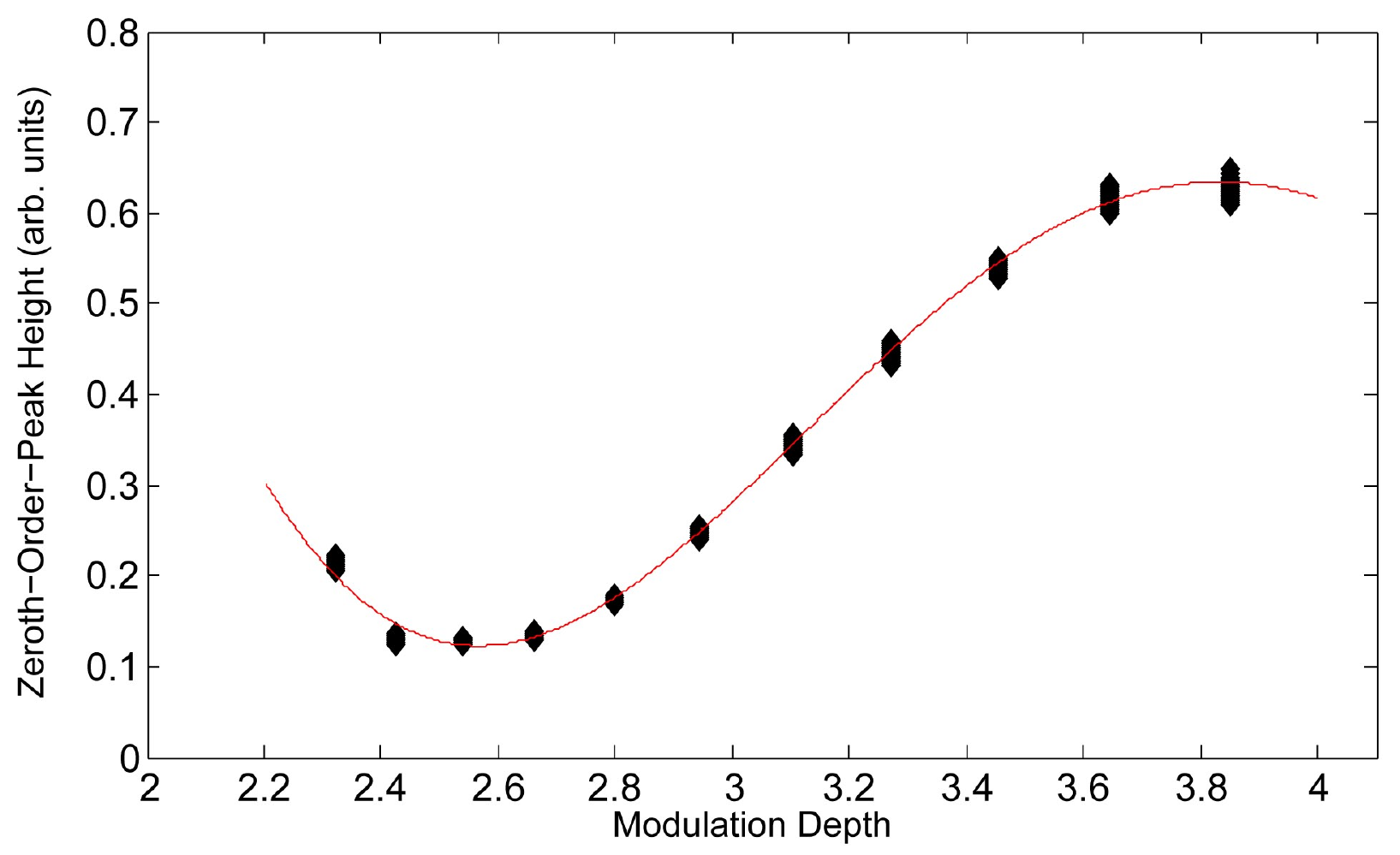}
\caption{A plot of zeroth order (carrier) peak height versus modulation depth. The solid red line is drawn to guide the eye.}
\label{fig:EOMP0}
\end{figure}

To explore the extent to which our experimental FM spectrum conforms to this prediction, we studied the intensity of the central carrier frequency, proportional to $J_0^2$, as we varied the  RF input power to the EOM, and thus the modulation depth. As Fig. \ref{fig:EOMP0} clearly indicates, we do not ever completely extinguish this carrier frequency component, whereas we would in the ideal case expect this frequency component  to vanish for $\beta \simeq 2.4$ (see Fig. \ref{Fig:BesselFcns}).  By considering similar plots of each of the first three sideband intensities over a range of EOM input RF powers, and using as reference points various values of RF power for which the graphs either minimize or cross one another, we can develop an approximate mapping of input power to modulation depth. Over our experimental range of $\beta$ values we find that expressing $\beta$ as a quadratic function of RF power gives an adequate mapping for present purposes.  This parametrization was used to generate the x-axis in Fig. \ref{fig:EOMP0}.

\begin{figure}[h]
\centering
\includegraphics[width = .5\textwidth]{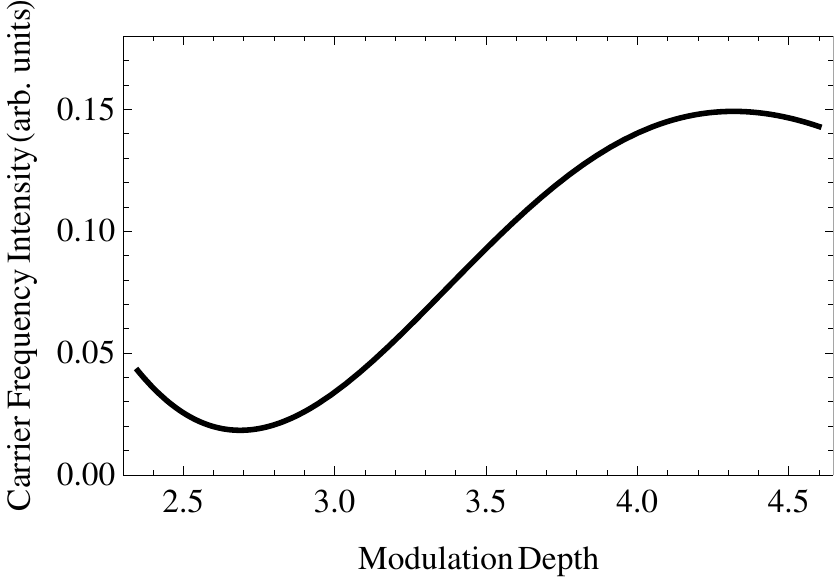}
\caption{A plot based on Eq. \eqref{Eq:DistMod} for $\lambda =0.8$. With this value for $\lambda$, expansion to second order is sufficient to reveal behavior. As we are not interested in quantitative predictions, higher numerical precision was not pursued. All of the qualitative information is contained in Eq. \eqref{Eq:DistMod}.
}
\label{Fig:DistMod}
\end{figure}

The simplest possible explanation of the non-ideality represented in Fig. \ref{fig:EOMP0}  is that a portion of the laser beam passing through the EOM somehow remains `unmodulated' (in this case, that portion would be roughly 2.5\% of the total laser intensity). A more physically realistic model for the observed behavior, which also explains the observed non-ideality in the higher-order sidebands, is to assume a spatially-dependent modulation strength within the electro-optic crystal of the EOM. Given the presence of the resonant RF cavity in which the EOM crystal is placed, it seems reasonable to assume that over the finite transverse extent of the laser beam, the light passing through the crystal will experience a range of RF powers. To test the reasonableness of this model, we can assign a modulation depth $\beta$ to the central region of the laser beam, while the outer annulus (assumed for simplicity to be of equal area) experiences a modulation depth $\lambda \beta$, where $\lambda<1$. Since the photodiode detector spatially averages the FM beam, the intensity of the $n^{th}$ sideband can be expressed as
\be
I_n  \propto \frac{1}{2}(J_n^2(\beta)+J_n^2(\lambda \beta)). \nonumber
\ee
Using a multiplication property of Bessel functions and assuming $\lambda$ is close to one, we find
\be
I_n \propto \frac{1}{2}J_n^2(\beta) + \frac{\lambda^{2n}}{2}\left(J_n(\beta)+\frac{1-\lambda^2}{2}\beta J_{n+1}(\beta) + \frac{(1-\lambda^2)^2}{8}\beta^2J_{n+2}(\beta)\right)^2.
\label{Eq:DistMod}
\ee
The predictions of this model are plotted in Fig. \ref{Fig:DistMod}. This simplistic model produces intensity vs. modulation depth results that at least qualitatively match those of Fig. \ref{fig:EOMP0}.  We find that the dependence of higher-order sideband intensities on modulation depth also deviate from the predictions of FM theory, and that this `spatially-distributed modulation' model can also qualitatively reproduce these non-idealities.

\section{Conclusions and ongoing work}

We have demonstrated the utility of high modulation depth frequency modulation spectroscopy in the context of high-precision atomic spectroscopy experiments. Careful modeling shows that the signal size is as much as an order of magnitude greater in the moderate modulation depth regime as compared to the more common low modulation depth regime. We reproduce the analytic FM profiles for arbitrary modulation depth with first harmonic demodulation, and extend this theory to second harmonic demodulation as well. We  have demonstrated excellent agreement between experimental data and these profiles in the regime where spectral line-widths are comparable in size to the modulation frequency.  We discuss and model several sources of deviation from ideal FM behavior and offer a simple model to explain these observations.

Given the very high resolution spectra we have produced with relatively low sample optical depths, and the success we have had in making high-precision measurements with these line shapes, we are moving forward to use FM techniques with our atomic beam apparatus to study other weak transitions.  We have undertaken a two-color, two-step atomic beam spectroscopy experiment in which the first laser is locked to the ground-state transition, and we scan the second-step laser across an excited state transition.  We expect this transition to exhibit an order-of-magnitude lower OD than the work described here.  Given the signal-to-noise ratio achieved already, we anticipate that FM spectroscopy of this transition should yield precise measurements of excited-state Stark shifts in indium\cite{SafMaj13}.  We would also like to study an intrinsically weak M1 transition in atomic thallium in our atomic beam environment.  This ground state $6p_{1/2} - 6p_{3/2}$ transition in thallium has been extensively studied in the high density vapor cell environment\cite{MajTsai}, and has important connections to parity non conservation work in this element\cite{Vetter}.  We estimate that in a thallium atomic beam, direct absorption measurements of this `forbidden' transition would yield  $10^{-5} < OD_{M1} < 10^{-4}$, making it a good candidate for exploration using the FM spectroscopy techniques employed here.

\section*{Acknowledgments}
We wish to acknowledge the support of the National Science Foundation RUI program through Grant No. 0969781.  We also acknowledge the contributions of Andy Schneider and Antonio Lorenzo at an earlier stage to the development of the atomic beam and Stark shift experiment.


\begin{thebibliography}{99}
\bibitem{Albrecht91} T. R. Albrecht, P. Grutter, D. Horne, and D. Rugar, ``Frequency modulation detection using high-Q cantilevers for enhanced force microscope sensitivity," J. Appl. Phys. {\bf 69,} 668 (1991).

\bibitem{Bjorklund80} G. C. Bjorklund, ``Frequency-modulation spectroscopy: a new method for measuring weak absorptions and dispersions," \ol {\bf 5,} 15 (1980).

\bibitem{Bjorklund81} G. C. Bjorklund and M. D. Levenson, ``Sub-Doppler frequency-modulation spectroscopy of ${\mathrm{I}}_{2}$," \pra {\bf 24,} 166 (1981).

\bibitem{BjorkFM} G. C. Bjorklund and M. D. Levenson, Appl. Phys. B, {\bf 32,} 145 (1983).

\bibitem{Grimm89} R. Grimm and J. Mlynek, ``Light-pressure-induced line-shape asymmetry of the saturation dip in an atomic gas," \prl {\bf63,} 232 (1989). 

\bibitem{Sansonetti95} C. J. Sansonetti, B. Richou, R. Engleman, and L. J. Radziemski, ``Measurements of the resonance lines of $^{6}\mathrm{Li}$ and $^{7}\mathrm{Li}$ by Doppler-free frequency-modulation spectroscopy," \pra {\bf 52,} 2682 (1995).

\bibitem{Vladimirova09} Y. V. Vladimirova, V. N. Zadkov, A. V. Akimov, A. Y. Samokotin, A. V. Sokolov, V. N. Sorokin, and N. N. Kolachevsky, ``Frequency-modulation spectroscopy of coherent dark resonances in 87Rb atoms," Appl. Phys. B {\bf 97,} 35 (2009).

\bibitem{Moerner86} W. E. Moerner, P. Pokrowsky, F. M. Schellenberg, and G. C. Bjorklund, ``Persistent spectral hole burning for R' color centers in LiF crystals: Statics, dynamics, and external-field effects," \prb {\bf 33,} 5702 (1986).

\bibitem{Moerner89} W. E. Moerner and L. Kador, ``Optical detection and spectroscopy of single molecules in a solid" \prl {\bf 62,} 2535 (1989).

\bibitem{Bomse92} D. S. Bomse, A. D. Stanton and J. A. Silver, ``Frequency modulation and wavelength modulation spectroscopies: comparison of experimental methods using a lead-salt diode laser," \ao {\bf 31,} 718 (1992).

\bibitem{Webster87} C. R. Webster and R. D. May, ``Simultaneous in situ measurements and diurnal variations of NO, NO2, O3, jNO2, CH4, H2O, and CO2 in the 40- to 26-km region using an open path tunable diode laser spectrometer," J. Geophys. Res. D {\bf 92,} 11931, (1987).

\bibitem{Sachse87} G. W. Sachse, G. F. Hill, L. O. Wade, and G. M. Perry, ``Fast-response, high-precision carbon monoxide sensor using a tunable diode laser absorption technique," J. Geophys. Res. D {\bf 92,} 2071, (1987).

\bibitem{Loewenstein88} M. Loewenstein, ``Diode laser harmonic spectroscopy applied to in situ measurements of atmospheric trace molecules," J. Quant. Spectrosc. Radiat. Transfer, {\bf 40,} 249 (1988).

\bibitem{Liu04} J. T. C. Liu, J. B. Jeffries, and R. K. Hanson, ``Large-Modulation-Depth 2f Spectroscopy with Diode Lasers for Rapid Temperature and Species Measurements in Gases with Blended and Broadened Spectra," \ao {\bf 43,} 6500 (2004).

\bibitem{Ranjit13} G. Ranjit, N. A. Schine, A. T. Lorenzo, A. E. Schneider, and P. K. Majumder, ``Measurement of the scalar polarizability within the 5${P}_{1/2}$-6${S}_{1/2}$ 410-nm transition in atomic indium," \pra {\bf 87,} 032506 (2013).

\bibitem{Saf09} M. S. Safronova, M. G. Kozlov, W. R. Johnson, and D. Jiang, ``Development of a configuration-interaction plus all-order method
	for atomic calculations," \pra {\bf 80,} 012516 (2009).

\bibitem{Saf13} M. S. Safronova, U. I. Safronova, and S. G. Porsev, ``Polarizabilities, Stark shifts, and lifetimes of the In atom," \pra {\bf 87,} 032513 (2013).

\bibitem{vetterth} P. M. Vetter, Ph.D. Thesis, Univ. of Washington, (1995).

\bibitem{ChiRei68} C. Chiarella and A. Reichel, ``On the evaluation of integrals related to the error function," Math. Comp {\bf 22,} 137 (1968).

\bibitem{Supplee94} J. M. Supplee, E. A. Shittaker, and W. Lenth, \ao {\bf 33,} 6294 (1994).

\bibitem{SafMaj13} M. S. Safronova and P. K. Majumder, ``Thallium 7$p$ lifetimes derived from experimental data and         \textit{ab initio}       calculations of scalar polarizabilities," \pra {\bf 87,} 042502 (2013).

\bibitem{MajTsai} P. K. Majumder and L. L. Tsai, ``Measurement of the electric quadrupole amplitude within the 1283-nm $6p_{1/2} - 6p_{3/2}$ transition in atomic thallium," \pra {\bf 60,} 267 (1999).

\bibitem{Vetter} P. A. Vetter, D. M. Meekhof, P. K. Majumder, S. K. Lamoreaux, and E. N. Fortson, ``Precise Test of Electroweak Theory from a New Measurement of Parity Nonconservation in Atomic Thallium," \prl {\bf 74,} 2658 (1995).

\end{thebibliography}
\end{document}